\documentclass[conference]{IEEEtran}
\IEEEoverridecommandlockouts
\usepackage{cite}
\usepackage{bm}
\usepackage{siunitx}
\usepackage{amsmath,amssymb,amsfonts}
\usepackage{stfloats}
\usepackage{graphicx}
\usepackage{textcomp}
\usepackage{xcolor}
\usepackage[linesnumbered, ruled, vlined]{algorithm2e}
\def\BibTeX{{\rm B\kern-.05em{\sc i\kern-.025em b}\kern-.08em
    T\kern-.1667em\lower.7ex\hbox{E}\kern-.125emX}}

\makeatletter
\newcommand{\linebreakand}{%
  \end{@IEEEauthorhalign}
  \hfill\mbox{}\par
  \mbox{}\hfill\begin{@IEEEauthorhalign}
}
\makeatother

\begin{document}

\title{Resonant Method-based Fully Automated Core Loss Measurement System for Sub-MHz Magnetics With Switched Capacitor Sequence\\

}

\author{
\IEEEauthorblockN{Haoyu Wang, Alex Hanson}
\IEEEauthorblockA{\textit{Department of Electrical and Computer Engineering} \\ \textit{University of Texas at Austin}\\ $\left\{\text{wanghaoyu, ajhanson} \right\}$@utexas.edu}
}

\maketitle
\begin{abstract}
Accurate loss characterization is essential for the design of high-frequency power magnetic components. State-of-the-art resonant characterization methods are attractive for high accuracy and low sensitivity, especially at the MHz regime. However, they predominantly rely on manual tuning and computationally intensive Fast Fourier Transform (FFT) analysis to identify resonant conditions, causing both inefficiencies and inaccuracies. To ensure accuracy and expedite the process, this paper proposes a fully automated measurement architecture, the core innovation of which lies in the integration of digitally-controlled switched capacitor sequences and onboard signal processing circuits,enabling automated sweeping of both frequency and drive level for complete and rapid characterization with no human intervention. A design guideline for the switched capacitor sequence is presented and common commercial electromechanical power relays are characterized to enable sub-MHz measurements. Experimental results for several different magnetic materials demonstrate that the proposed system has great accuracy and is able to collect more than 1000 data points within 20 seconds, providing a very fast and robust solution for high-frequency magnetic characterization.

\end{abstract}


\section{Introduction}
With the increasing need for high power density and capability in modern power electronics applications \cite{hwang_ups, hwang_cps, hwang_zvs, dmou1, dmou2, jzheng1, jzheng2, jzheng3, jzheng4}, magnetic components deployed across a wide spectrum of frequencies and power levels have become a critical bottleneck in energy conversion \cite{CSullivan1, CMathuna}. Due to the significant contribution that magnetic components have to the overall physical volume and power dissipation of power converters, there is a strong trend to elevate operating frequencies (i.e., switching frequencies) to the MHz regime where the loss characterization of magnetics is critical and difficult. Therefore, precisely characterizing the loss metrics of magnetic components within a high-frequency spectrum has become increasingly important for both design optimization and performance evaluation. 

Generally, the power loss in magnetic components is categorized into copper loss (also known as winding loss) and core loss. To date, copper loss can be accurately evaluated by theoretical analysis or finite-element-analysis simulations \cite{yhan_copper, CSullivan2}. However, the precise characterization of core loss remains a formidable challenge across diverse drive levels and frequencies due to the inherent non-linearities of magnetic materials. The estimation of core loss typically relies on empirical models, such as Steinmetz equations \cite{CSteinmetz, kvenkatachalam} and modified ones \cite{jmuhlethaler, JReinert}. However, these empirical methods only yield rough approximations and generally require speculative extrapolation or extrapolation beyond the available data. Consequently, it is necessary to experimentally characterize raw magnetic materials and assembled magnetic components, thus validating their operational feasibility in real applications.

The effective loss of magnetic components is typically modeled as an equivalent series or parallel resistance that changes substantially with frequency and drive level. Small-signal instruments, such as conventional LCR meters or impedance analyzers, are inadequate for wide-range loss evaluation because small-signal parameters cannot be effectively extrapolated to large-signal operating conditions. In the literature, the most intuitive approach large-signal is the direct integration of the instantaneous voltage and current (i.e., $\int I \times V dt$) through the magnetic component \cite{mmu}. The direct $I \times V$ method is convenient and flexible under different shapes of waveforms (e.g., sinusoidal, square, and trapezoidal; with or without DC bias) \cite{jmuhlethaler2}. However, it suffers from severe inaccuracies with increasing frequency, especially in the MHz regime, due to the inevitable phase mismatch between the voltage and current sensing channels \cite{mmu}. An effective alternative is the calorimetry method that can provide high accuracy, but it demands complex experimental setups and remains susceptible to significant environmental thermal errors \cite{ppapamanolis}. 

To address these limitations, resonant measurement methods have emerged as an effective approach to evaluating magnetic materials and components from low to very high frequencies (i.e., beyond MHz level). Specifically, a series-resonant method has been proposed in \cite{yhan} and modified in \cite{mmu}, which cancels the inductance of the Device Under Test (DUT) using a series capacitance that resonates at the operating frequency so that the equivalent resistance of the DUT (i.e., the total power loss) can be easily measured at any drive levels. However, the equivalent resistance is usually small and the compensated resonant network generally behaves as a low impedance to the source, which is prone to voltage sensing errors. Therefore, a parallel-resonant method has been proposed in \cite{abrown}, which utilizes a parallel resonant capacitance to elevate the equivalent impedance of the network and presents opportunities to incorporate dc bias. These resonant techniques are both capable of evaluating core loss with high accuracy in the MHz regime due to the precise reactance cancellation.

Nevertheless, resonant approaches are inherently labor-intensive, time-consuming and error-prone. In practice, the operator first must manually solder resonant capacitors to change the resonant frequency, fine-tunes the frequency to precisely operate at resonance, capture the voltage and/or current magnitudes, and subsequently derive the core loss. This process is further complicated by the non-sinusoidal waveforms from power amplifiers (PAs) \cite{abrown} or square waves from switching circuits \cite{janderson_classd}. In such scenarios, extracting the magnitude and phase of the fundamental-frequency components is demanded prior to tuning and measurement, which typically requires the FFT function on an advanced oscilloscope or computing powers from higher-level computational tools. Furthermore, the slow process of manual tuning makes it extremely difficult to maintain a controlled temperature for DUT, as the magnetic component experiences self-heating during the prolonged tuning and measurement cycle. Consequently, comprehensive resonant measurements across temperature variations, or even at a stable controlled temperature, have remained largely unattempted in the literature \cite{nkirkby}.

To expedite data acquisition and reduce thermal drift, it is imperative to develop high-speed automated loss measurement systems for resonant methods. Automated platforms can enable the fast collection of extensive datasets of core loss, which is essential to improve the accuracy of traditional empirical models and train emerging machine learning-based core loss models \cite{dserrano}. Moreover, rapid measurement ensures negligible self-heating in the DUT, thereby keeping material properties relatively constant during the test. While automated systems are commercially available for direct $I \times V$ measurements at frequencies up to hundreds of kilohertz, automation for resonant methods within the MHz regime remains a big challenge. In the literature, an automated system utilizing the series-resonant approach was proposed in \cite{janderson}, which relies on stepper motors for frequency sweeping, real-time waveform acquisition by digital oscilloscopes, and subsequent FFT analysis in MATLAB. This architecture is heavily equipment-dependent and computationally intensive, which is effective but inefficient. To expedite the evaluation process, \cite{hwang} utilized high-frequency analog conditioning circuits \cite{hwang2, hwang3} to extract the fundamental components of signals of interest in resonant methods, thus eliminating the involvement of oscilloscopes and MATLAB. However, it lacks the capability of frequency sweeping due to the lack of the resonant capacitor swapping system.  

In this paper, a fully automated loss measurement system is proposed for resonant methods to improve both the quantity and quality of measured data, which operates at sub-MHz but in principle can be scaled to tens of MHz, far surpassing the reasonable limits of direct $I \times V$ approaches. Specifically, it incorporates switched capacitor sequences for frequency sweeping and can evaluate the loss characterization in a fully automated way without any human supervision and FFT analysis. While commercial electromechanical power relays are chosen for sub-MHz operation in the proposed system, the characterization of the switches in the sequences shows potentials for very-high-frequency opportunities. The rest of the paper is organized as follows. Section \ref{sec2} introduces the operation principles of the proposed automated system. Section \ref{sec3} characterizes the switches in the switched capacitor sequences presents a design guideline. Section \ref{sec4} verifies the feasibility and superiority of the proposed automated system with experimental results. Section \ref{sec5} concludes the paper.

\section{Operation Principles of the Proposed Automated System}\label{sec2}

\begin{figure*}[tp]
\centering{\includegraphics[width=18cm]{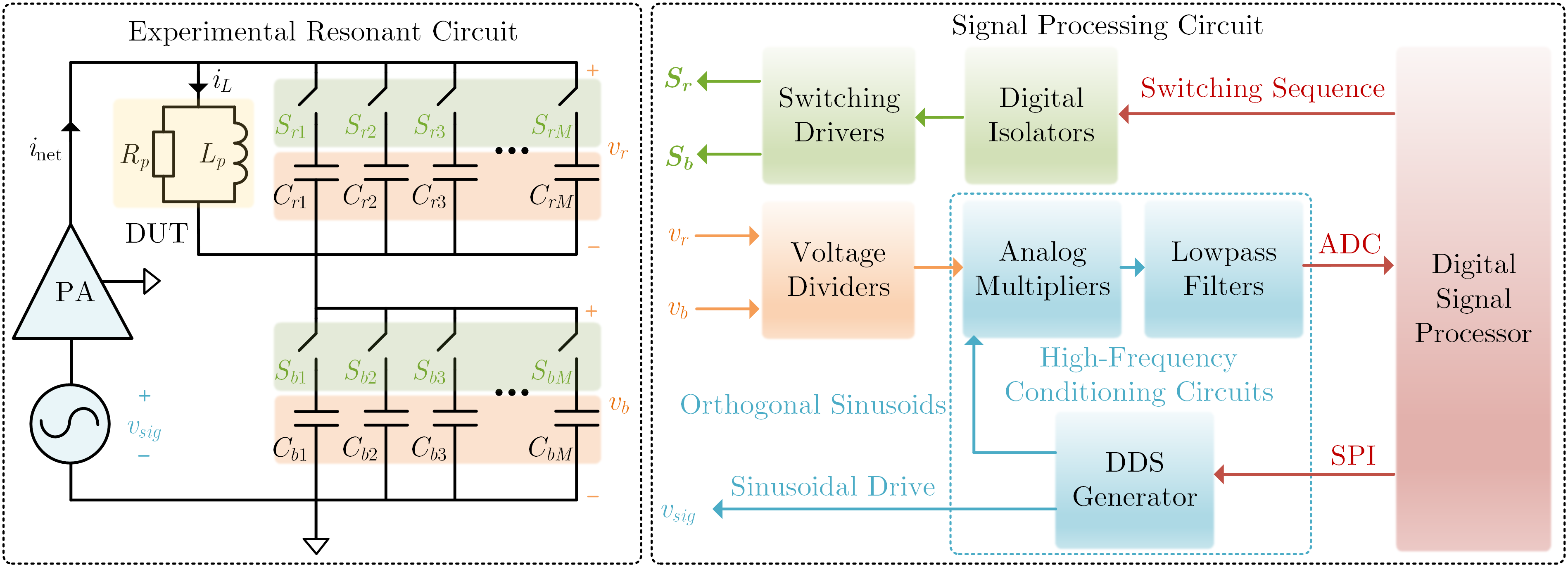}}
\caption{The proposed automated loss measurement system with switched capacitor sequences, high-frequency conditioning circuits and local microcontroller.}
\label{fig:system}
\end{figure*} 

\subsection{Experimental Resonant Circuit}

The proposed automation framework shown in Fig.~\ref{fig:system} can be generally applied to various resonant measurement methods that cancel the reactance of the DUT by a tuning capacitor, thereby isolating the measurement of the equivalent resistance at a single frequency and drive level. These methods rely inherently on sinusoidal drive and measurement, although harmonics may be introduced intentionally or inevitably. During the measurement process, the compensating resonant capacitance is roughly tuned at a specified resonant frequency and the drive magnitude is swept near the specified resonant frequency. The operating frequency is fine-tuned at each drive magnitude to guarantee precise operation within a narrow frequency variation. Subsequently, this process is repeated across a wide range of resonant frequencies, which requires replacement of the resonant capacitor. 

Take the parallel resonant method for illustration, as shown in Fig.~\ref{fig:system}. The entire resonant circuit is excited by the output of a PA driven by a digitally-controlled signal $v_{sig}$ at a designated drive level. The DUT can be regarded as an inductor $L_p$ in parallel with an effective resistor $R_p$ that represents the drive-level-dependent core loss or the total loss if the winding loss is significant. A switched resonant capacitor sequence $C_{r1}\text{-}C_{rM}$ is installed to resonate with the parallel-equivalent reactance of the DUT, leaving the impedance of the resonant network purely resistive with a voltage $v_r$. A switched blocking capacitor sequence $C_{b1}\text{-}C_{bM}$ with a voltage $v_b$ is also placed to measure the net current $i_{\text{net}}$. $v_r$ and $v_b$ are measured and provide sufficient information to calculate $R_p$. The switched capacitor sequences are intended to be in binary form, although available capacitor values deviate from a precise binary sequence and the actual capacitor values must be used in calculations. Note that the switched blocking capacitor sequence can be replaced with any current sensor with appropriate modifications. Specifically, non-capacitive current sensing can potentially permit DC current through the DUT, offering important insight into DC-biased loss characteristics.


 \subsection{Core Loss Calculation}

Assuming a purely sinusoidal output from the PA, all signals within the resonant circuit are considered sinusoidal without higher-order harmonics. Under this ideal condition, a precise $90^\circ$ phase difference exists between $v_r$ and $v_b$ when $C_r$ achieves resonance with $L_p$. The phase relationship serves as a definitive indicator for identifying the resonance when the frequency is fine-tuned at a given capacitance and drive level. At resonance, the magnitudes of $v_r$ and $v_b$ (denoted as $V_r$ and $V_b$, respectively) are detected, and the corresponding fine-tuned resonant frequency $f_r$ is recorded. These experimental measurements are subsequently used to evaluate the core loss characteristics.

Based on the fine-tuned resonant frequency, the effective inductance of the DUT is given by:
\begin{equation} 
\label{eq:L_eff}
L_{\text{eff}} = \frac{1}{(2\pi f_r)^2 C_r}.
\end{equation}

The root-mean-square (RMS) values of the net current $I_{\text{net,rms}}$ and the inductor current $I_{L,\text{rms}}$ are expressed as:
\begin{equation} 
\label{eq:I_net_rms}
I_{\text{net,rms}} = \sqrt{2}\pi f_r C_b V_b.
\end{equation}
\begin{equation} 
\label{eq:I_L_rms}
I_{L,\text{rms}} = \pi f_r \sqrt{2C_b^2 V_b^2 + 2C_r^2 V_r^2}.
\end{equation}

Accordingly, the parallel-equivalent resistance of the DUT can be derived as:
\begin{equation} 
\label{eq:R_p}
R_p = \frac{V_{r,\text{rms}}}{I_{\text{net,rms}}} = \frac{V_r}{2\pi f_r C_b V_b}.
\end{equation}

The quality factor ($Q$) of the DUT is given by:
\begin{equation} 
\label{eq:Q_factor}
Q = \frac{R_p}{2\pi f_r L_{\text{eff}}} = \frac{C_r V_r}{C_b V_b}
\end{equation}

Consequently, the overall volumetric power loss of the DUT at the specific operating point is calculated by:
\begin{equation} 
\label{eq:loss}
P_v = \frac{V_{r,\text{rms}}^2}{R_p V_{\text{eff}}} = \frac{\pi f_r C_b V_b V_r}{V_{\text{eff}}}
\end{equation}
where $V_{\text{eff}}$ denotes the effective volume of the DUT. Furthermore, the peak flux density $B_{\text{pk}}$ can be quantified as follows:
\begin{equation} 
\label{eq:Bpk}
B_{\text{pk}} = \frac{\sqrt{2}L_{\text{eff}} I_{L,\text{rms}}}{N A_{\text{eff}}} = \frac{\sqrt{C_b^2 V_b^2 + C_r^2 V_r^2}}{2\pi f_r C_r N A_{\text{eff}}}
\end{equation}
where $N$ represents the number of winding turns and $A_{\text{eff}}$ is the effective cross-sectional area of the DUT.

\begin{figure*}[tp]
\centering{\includegraphics[width=18cm]{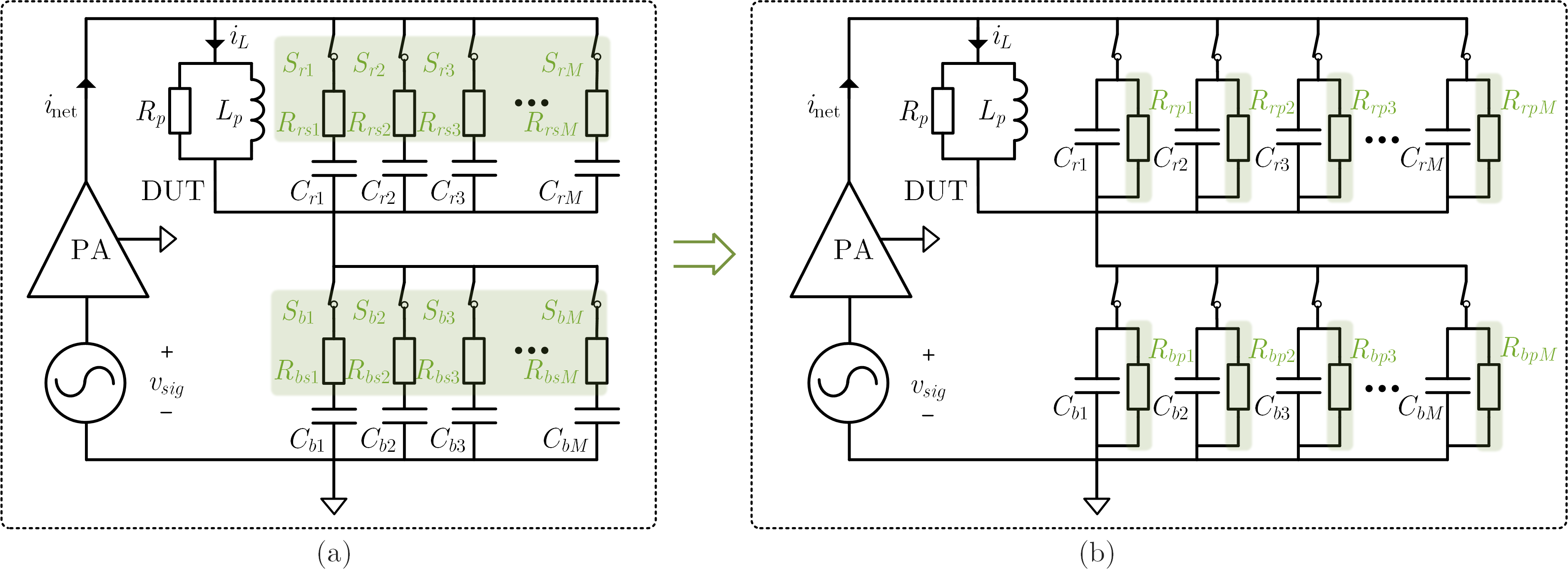}}
\caption{The equivalent circuit diagrams showing the impacts of the on-state resistance of the digitally-controllable switches. (a) Series-equivalent circuit. (b) Parallel-equivalent circuit.}
\label{fig:onstate_resistance}
\end{figure*}

\subsection{Signal Processing Circuit}

Although the driving signal $v_{\text{sig}}$ is typically sinusoidal, harmonic components are inevitably introduced by the non-linearity of the PA, particularly when the impedance of the overall power circuit deviates significantly from the nominal load impedance. In particular, substantial harmonic distortion may appear in the blocking capacitor voltage within the parallel-resonant configuration. It impedes the sensing of the fundamental phases, the precise identification of the resonance, and the measurement of the fundamental amplitudes, thereby introducing significant errors into the core loss evaluation. Therefore, it is imperative to extract the fundamental components of $v_r$ and especially $v_b$ at resonance to validate core loss calculations.

The conventional approach to eliminating harmonic interference relies on FFT algorithms implemented in digital oscilloscopes or post-processing software such as MATLAB. These methods substantially prolong the measurement cycle and become equipment-specific, which is undesirable for fully automated systems. As a high-speed equipment-agnostic alternative, the high-frequency conditioning circuits proposed in \cite{hwang} can be utilized to achieve rapid harmonic isolation, as illustrated in Fig.~\ref{fig:system}. The conditioning circuits are implemented in an analog way to extract the fundamental magnitudes and phases of $v_r$ and $v_b$, enabling ultra-fast signal processing compared to digital FFT implementations.

The operating principle of the conditioning circuits is detailed as follows. A Direct Digital Synthesis (DDS) generator outputs: 1) a driving signal $v_{\text{sig}}(t)$ to a radio-frequency PA with adjustable frequency and magnitude, and 2) two orthogonal reference sinusoids, $\sin\omega t$ and $\cos\omega t$ with the same frequency as $v_{\text{sig}}$. The PA subsequently amplifies $v_{\text{sig}}$ to excite the resonant tank. The orthogonal reference signals are then employed 
to extract the fundamental magnitudes and phases of $v_r(t)$ and $v_b(t)$, successfully isolating the higher-order harmonics generated by the PA.

Taking $v_r(t)$ as an example, this periodic signal typically exhibits zero DC bias when the net current sensing is implemented capacitively. It can be expressed as a Fourier series by $v_r(\omega t) = \sum_{n=1}^{\infty} V_{r,n} \sin(n\omega t + \varphi_{r,n})$, where $\omega=2\pi f$ denotes the fundamental angular frequency, while $V_{r,n}$ and $\varphi_{r,n}$ represent the magnitude and phase of the $n$-th order components, respectively. When $v_r(\omega t)$ is sensed through a voltage divider with a scaling gain $k_r$ and subsequently multiplied by $\sin\omega t$ and $\cos\omega t$ through analog multipliers, it yields:

\begin{subequations}
\label{eq:harmonics_sub}
\begin{align}
&k_r v_r(\omega t) \cdot \sin(\omega t) = \sum_{n=1}^{\infty} \frac{k_r V_{r,n}}{2} \Big\{ \nonumber \\
&\quad \cos\left[ (n-1)\omega t + \varphi_{r,n} \right] - \cos\left[ (n+1)\omega t + \varphi_{r,n} \right] \Big\}. \label{eq:harmonics_sin} \\[1.5ex]
&k_r v_r(\omega t) \cdot \cos(\omega t) = \sum_{n=1}^{\infty} \frac{k_r V_{r,n}}{2} \Big\{ \nonumber \\
&\quad \sin\left[ (n-1)\omega t + \varphi_{r,n} \right] + \sin\left[ (n+1)\omega t + \varphi_{r,n} \right] \Big\}. \label{eq:harmonics_cos}
\end{align}
\end{subequations}

The resulting product signals go through low-pass filters (LPFs), leaving the DC components that preserve the fundamental components of $v_r(\omega t)$:
\begin{equation}
V_{r,d} = \frac{k_r V_{r,1}}{2}\cos\varphi_{r,1}, \quad V_{r,q} = \frac{k_r V_{r,1}}{2}\sin\varphi_{r,1}.
\end{equation}

The isolated DC components are sampled by Analog-to-Digital Converters (ADCs) of a local Digital Signal Processor (DSP). The DSP extracts the fundamental phase and magnitude of $v_r(\omega t)$ through the following operations:
\begin{equation}
\varphi_{r,1} = \arctan(V_{r,q}/V_{r,d}), \quad V_{r,1} = \frac{2\sqrt{V_{r,d}^2 + V_{r,q}^2}}{k_r}.
\end{equation}

The fundamental components of $v_{b}$ (i.e., phase $\varphi_{b,1}$ and magnitude $V_{b,1}$) are extracted in the same way. Subsequently, the DSP determines the phase difference $\varphi = \varphi_{r,1} - \varphi_{b,1}$, fine-tunes the excitation frequency to establish resonance for each specified drive level, and captures $V_{r,1}$ and $V_{b,1}$ at resonance. Finally, the volumetric core loss defined in \eqref{eq:loss} and the peak flux density in \eqref{eq:Bpk} are refined as:
\begin{equation}
P_v = \frac{\pi f_r C_b V_{b,1} V_{r,1}}{V_{\text{eff}}}, \quad B_{\text{pk}} = \frac{\sqrt{C_b^2 V_{b,1}^2 + C_r^2 V_{r,1}^2}}{2\pi f_r C_r N A_{\text{eff}}}.
\end{equation}




\section{Analysis and Implementation of Switches in Switching Sequences}\label{sec3}

\begin{figure*}[tp]
\centering{\includegraphics[width=18cm]{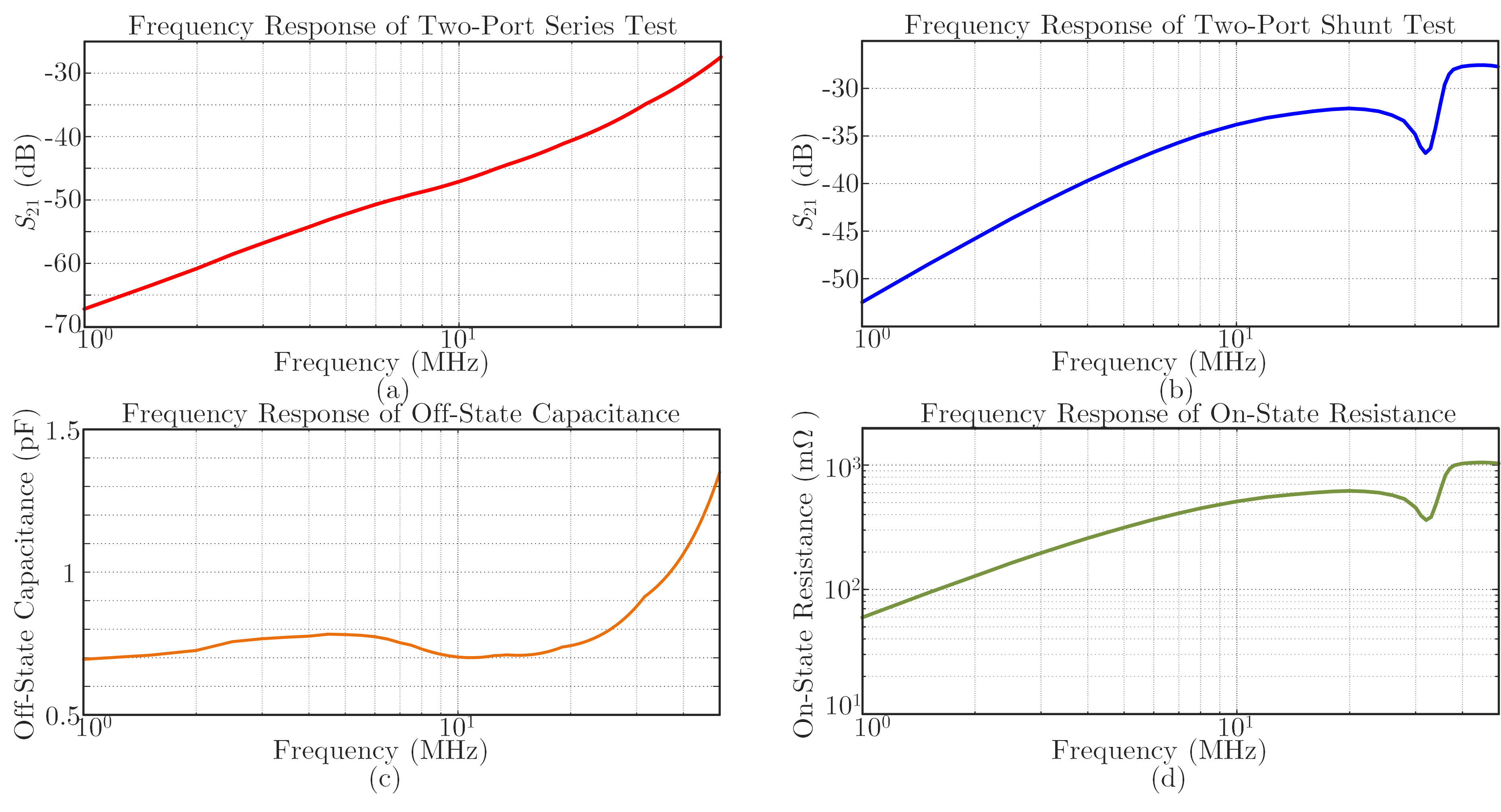}}
\caption{Experimental characterization of the selected T77V1D10-12 power relay across the MHz-level frequency spectrum. (a) $S_{21}$ transmission characteristics in off-state two-port series test. (a) $S_{21}$ transmission characteristics in on-state two-port shunt test. (c) Extracted $C_{\text{off}}$. (d) Extracted $R_{\text{on}}$. }
\label{fig:switch}
\end{figure*}

Switches integrated into the two capacitor sequences are fundamental to enabling automated frequency sweeping. This section delves into the analytical evaluation of switch-induced non-idealities on the system-level core loss measurement accuracy. Furthermore, a systematic test and selection guideline for the switching devices are formulated.

\subsection{Switch-Induced Non-Idealities}

In practical applications, switches exhibit non-ideal characteristics, inherently maintaining parasitics across different switching states, namely on-state resistance and off-state capacitance. These parasitics become increasingly significant as the operating frequency escalates, thereby requiring rigorous consideration during system design.

Note that the primary objective of resonant measurement methodologies is to characterize the equivalent resistance of the DUT once its reactance is fully compensated for. However, when the switches are conducting, they inevitably introduce an on-state resistance in series with the designated capacitance network. It is particularly important within the resonant capacitance 
branches, potentially introducing non-negligible measurement errors into the extracted DUT resistance.

As illustrated in Fig.~\ref{fig:onstate_resistance}(a), denote $R_{rsk}$ as the on-state resistance of the switch $S_{rk}$ within the resonant switching sequence, which appears in series with the corresponding resonant capacitance component $C_{rk}$. For a parallel-resonant configuration, this series combination can be transformed into an equivalent parallel network, comprising the capacitance $C_{rk}$ and an equivalent parallel resistance $R_{r\text{p}k}$, as shown in Fig.~\ref{fig:onstate_resistance}(b). The transformation satisfies the following relationship:
\begin{equation}
\label{eq:resistance_approx}
R_{r\text{p}k} = \frac{1}{\left(\omega C_{rk} R_{r\text{s}k}\right)^2 + 1} R_{r\text{s}k} \approx \frac{1}{\omega^2 C_{rk}^2 R_{r\text{s}k}}
\end{equation}

Consequently, each branch in the resonant switching sequence containing an on-state switch introduces an equivalent resistance that appears in parallel with the parallel-equivalent core loss resistance $R_p$ of the DUT. To ensure the measurement accuracy of the DUT resistance, the condition $R_{r\text{p}k} \gg R_p$ must be strictly satisfied. Based on the approximation in \eqref{eq:resistance_approx}, this condition implies a stringent constraint on the maximum on-state resistance:
\begin{equation}
\label{eq:resistance_constraint}
R_{r\text{s}k} \ll \frac{1}{\omega^2 C_{rk}^2 R_p}
\end{equation}

Accordingly, minimizing the on-state resistance $R_{\text{on}}$ is imperative for precise DUT resistance measurements. Furthermore, to reduce interference from off-state switches on the designated capacitance branches (especially the first branch), a low off-state capacitance $C_{\text{off}}$ is stringently required, which indicates that minimizing the switch Figure of Merit (FOM), i.e., $R_{\text{on}}C_{\text{off}}$, is a primary prerequisite for device selection.

\subsection{Test and Implementation of Switches}

Based on the above analysis, the switches deployed in the switching sequences must satisfy three requirements: 1) digital controllability, 2) high power capacity, and 3) a minimized FOM. A comparative evaluation of available switching devices highlights their respective limitations: manual mechanical switches lack digital control interfaces; micro-electro-mechanical-system relays exhibit insufficient power capacity; and semiconductor devices typically have excessively large FOMs that even match those of the resonant tank. Therefore, electromechanical power relays represent an ideal solution for the automated loss measurement system, successfully meeting all design specifications.

Although the commercial market offers a wide array of electromechanical power relays, their manufacturer datasheets frequently omit FOM characteristics, which requires a rigorous pre-selection experimental evaluation. In this study, practical two-port series and shunt tests were developed to measure $C_{\text{off}}$ and $R_{\text{on}}$ of the switches, respectively, using a high-precision impedance analyzer. An evaluation board was constructed to characterize the FOMs of multiple commercial power relays. Consequently, T77V1D10-12 relay was selected as the switching device for the proposed automated system, as it exhibited the optimal FOM among all candidate switches. 

Experimental results are illustrated in Fig.~\ref{fig:switch}, where $S_{21}$ represents the transmission characteristic from two-port measurements. The selected switch maintains a relatively low $C_{\text{off}}$ of approximately \SI{1}{pF} across the \num{1}-\SI{50}{MHz} spectrum, which is entirely negligible compared to the minimum branch capacitance (typically on the order of hundreds of pF). Conversely, $R_{\text{on}}$ scales significantly with operating frequency due to high-frequency skin and proximity effects, demonstrating that the selected device is only feasible for frequencies up to \SI{1}{MHz}. Consequently, the proposed automated system is designed for the sub-MHz regime, but it establishes a scalable framework with substantial potential for higher MHz operation, pending the availability of advanced switching devices. 

\section{Experimental Implementation and Validation}\label{sec4}

\subsection{Experimental Setup and Implementation}

\begin{figure}[t]
\centerline{\includegraphics[width=9cm]{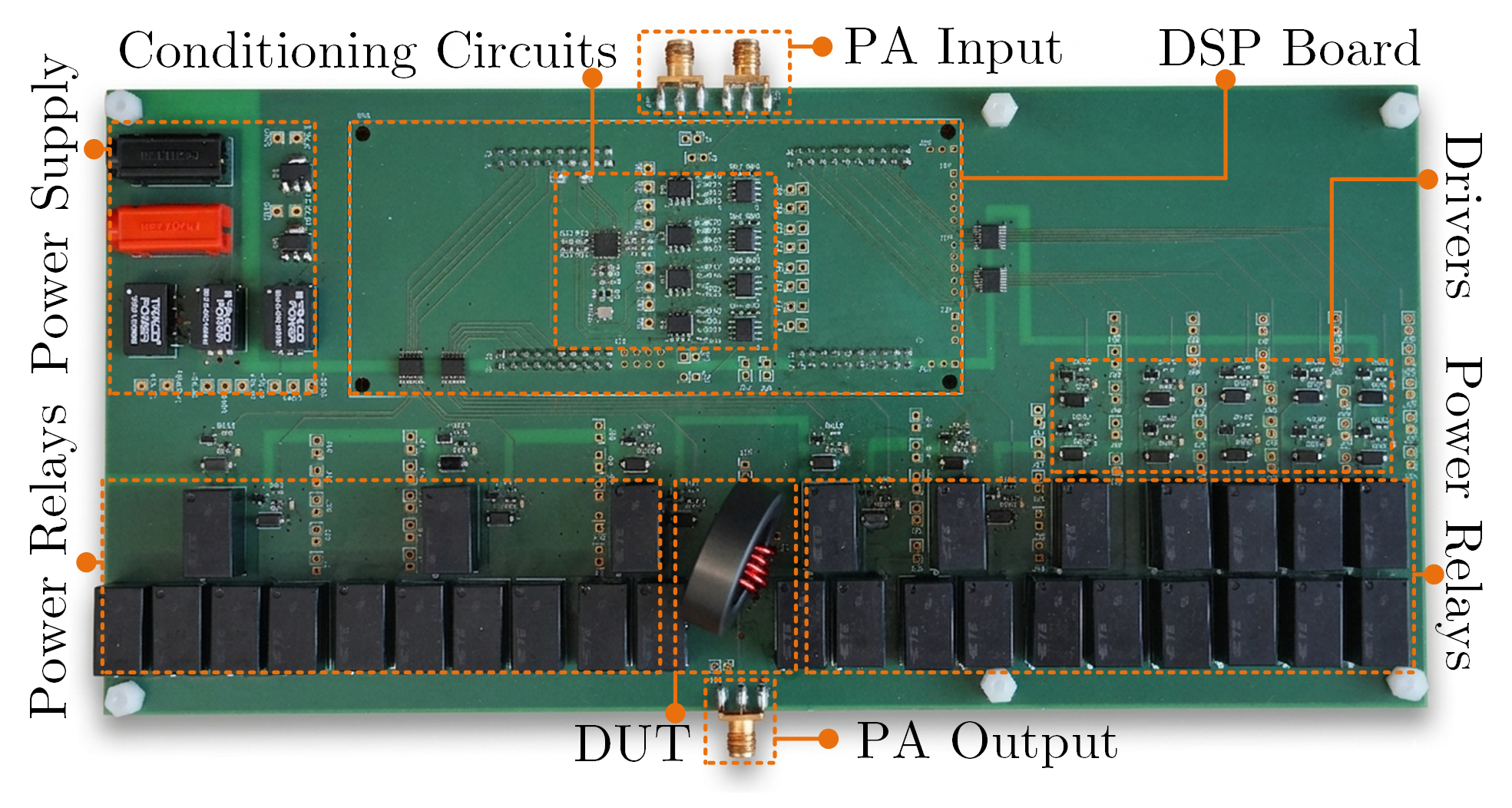}}
\caption{The PCB diagram of the proposed fully automated loss measurement system.}
\label{fig:board}
\end{figure} 

To validate the proposed measurement methodology, a fully automated loss measurement prototype integrating the high-frequency conditioning circuits and the selected T77V1D10-12 electromechanical power relays was developed, as illustrated in Fig.~\ref{fig:board}. The hardware architecture consists of a measurement motherboard stacked vertically above a digital control subsystem based on the TMS320F28379D DSP LaunchPad. The local DSP commands an AD9106 DDS generator through the Serial Peripheral Interface (SPI) protocol to synthesize the high-precision driving signal and orthogonal reference sinusoids. Concurrently, it samples the isolated DC components using its embedded high-resolution ADCs and executes an automated frequency-and-drive sweeping algorithm for automated core loss data acquisition \cite{hwang3}.  

Each of the dual-sequence switching sequences comprises 10 independent capacitance branches, theoretically enabling $2^{10}$ distinct resonant frequency configurations. To further mitigate switch-induced non-idealities, multiple power relays are 
paralleled in resonant branches with larger capacitances, thereby reducing the effective on-state resistance. The complete experimental setup is driven by an AR 100A400A radio-frequency PA with a wide operating bandwidth of 0.01-400 MHz. Furthermore, the DUT is positioned in the geometric center of the switching sequences to minimize stray parasitics.

\subsection{Accuracy Validation of the Proposed System}\label{}

\begin{figure*}[tp]
\centering{\includegraphics[width=18cm]{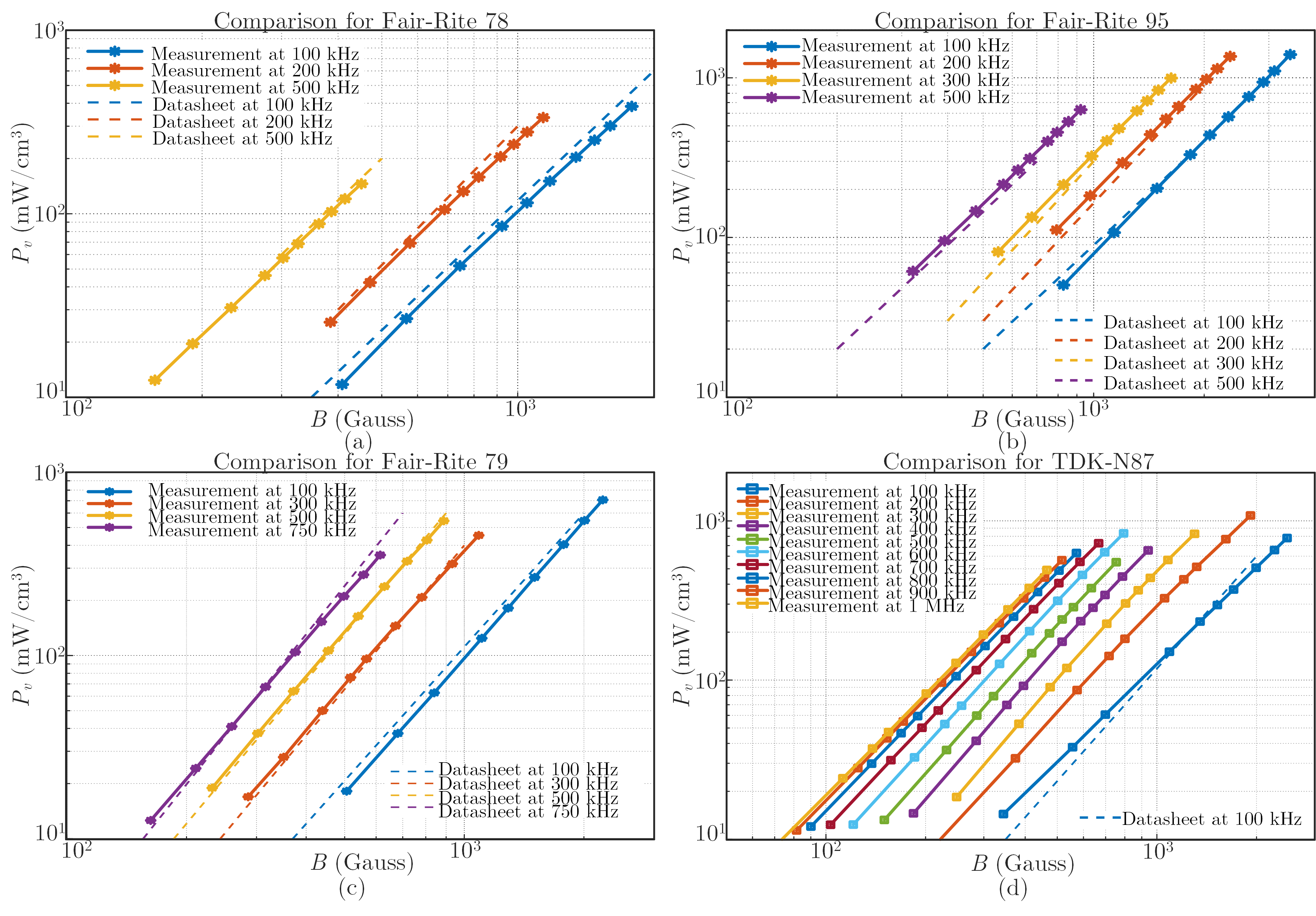}}
\caption{Experimental core loss measurement results and comparisons with manufacturer datasheets. (a) Fair-Rite 78. (b) Fair-Rite 95. (c) Fair-Rite 79. (d) TKD-N87. }
\label{fig:accuracy}
\end{figure*}

Measurement accuracy stands as a primary performance metric for the developed automated core loss characterization system. To rigorously validate the feasibility and accuracy of the proposed system, multiple distinct magnetic materials were extensively evaluated using the experimental prototype across wide spectrum of frequencies. The number of winding turns for DUT cores is restricted to either four or five, which minimizes the copper volume and winding resistance, thereby makingg the associated winding loss negligible relative to the core loss. The experimental test results are compared against the baseline data provided in the manufacturer datasheets to quantify the measurement fidelity.

The experimental validation incorporates a diverse set of manganese-zinc (MnZn) ferrite materials spanning both the Fair-Rite and TDK product family to verify the system performance across highly distinct frequency bands. Specifically. Four corresponding toroidal cores were chosen as DUT components: 
1) Fair-Rite 78 material (Part No. 5978001701, $\mu_i = 2300$) typically below \SI{200}{kHz}; 
2) Fair-Rite 95 material (Part No. 5995001801, $\mu_i = 3000$) from \SI{25}{kHz} to \SI{500}{kHz}; 
3) Fair-Rite 79 material (Part No. 5979001801, $\mu_i = 2300$) from \SI{500}{kHz} up to \SI{1}{MHz}; and 
4) TDK N87 material (Part No. B64290L0652X087, $\mu_i = 2200$) up to \SI{1}{MHz}. 

The experimental core loss measurement results for the four magnetic materials at \SI{25}{\celsius} are systematically illustrated in Fig.~\ref{fig:accuracy}, where the calculated volumetric core loss ($P_v$) is plotted against the peak flux density ($B_{\text{pk}}$) on a log-log scale. As depicted in Figs.~\ref{fig:accuracy}(a)--(c), the proposed automated system 
rigorously evaluates the Fair-Rite 78, 95, 79, and TDK N87 materials at multiple representative frequencies up to \SI{750}{kHz}. The measured loss curves (solid lines with markers) exhibit an exceptionally tight overlap with the baseline curves extracted from the manufacturer datasheets (dashed lines). Quantitatively, the maximum percentage error with respect to the datasheet specifications is capped within \num{13}\% and the vast majority of the captured experimental data points fall well within a tight $\pm$\num{5}\% error margin. This comprehensive multi-material evaluation across a wide flux range (from \SI{100}{Gauss} to over \SI{2000}{Gauss}) and a wide frequency range (from \SI{100}{kHz} to \SI{1}{MHz}) effectively validates the high fidelity and robustness of the proposed automated measurement system across the designated sub-MHz spectrum.

\subsection{Data Acquisition Speed Test}\label{}

\begin{figure*}[tp]
\centering{\includegraphics[width=18cm]{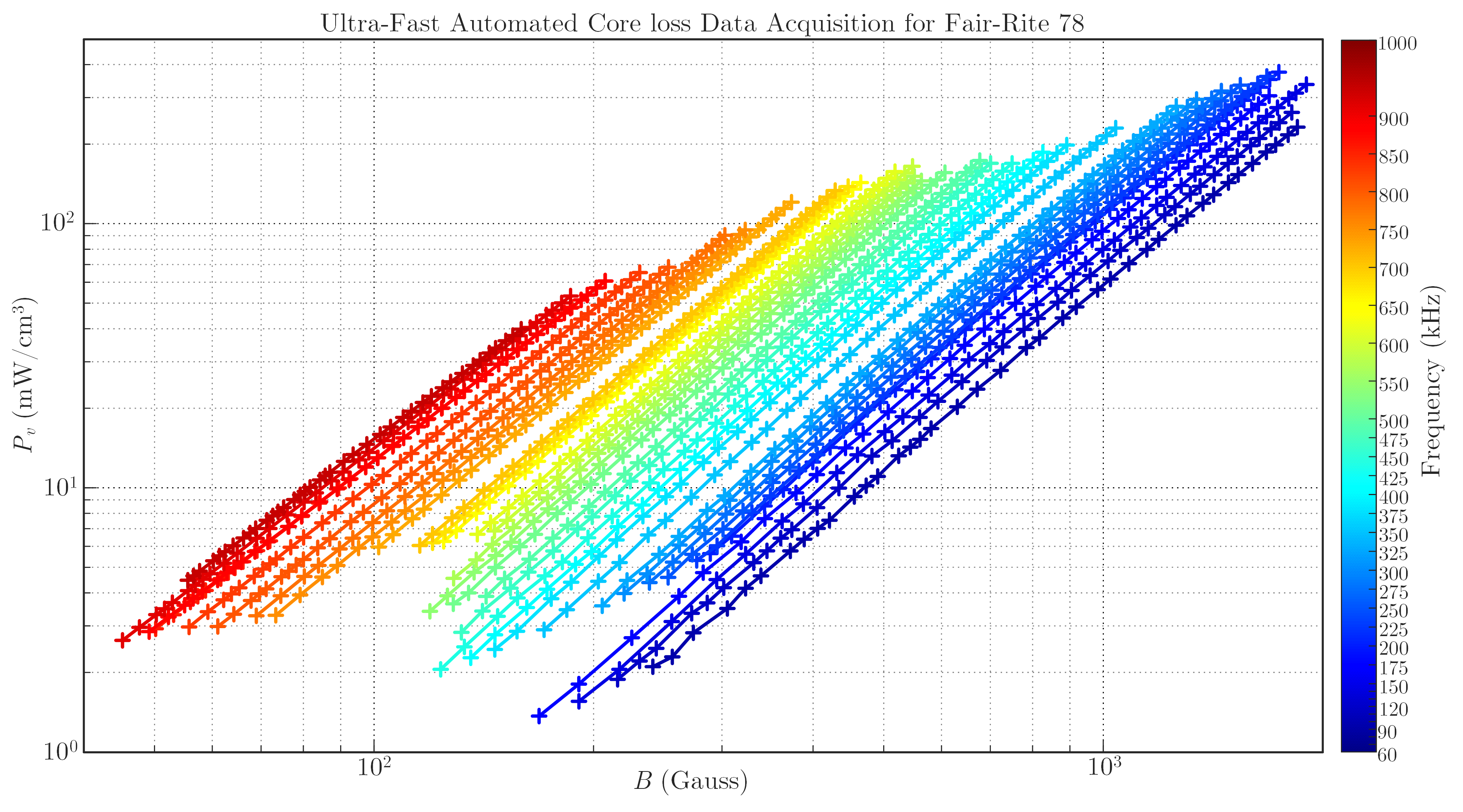}}
\caption{High-throughput core loss data acquisition for Fair-Rite 78 using the proposed automated measurement system. }
\label{fig:speed}
\end{figure*}

Another critical performance metric for the automated measurement system is its high-throughput data acquisition speed. To evaluate this capability, a comprehensive frequency sweep was executed across the full operational range of the Fair-Rite 78 
material from \SI{60}{kHz} up to \SI{1}{MHz}, comprising \num{33} frequencies with \num{40} data points evaluated at each frequency step. As illustrated in Fig.~\ref{fig:speed}, the proposed system successfully collected a total of \num{1320} discrete experimental data points within a time of only \SI{20}{seconds}. This translates to a \num{4000}$\times$ speed acceleration compared to conventional manual measurements performed by a proficient operator (typically restricted to \num{1} point per minute). Furthermore, it achieves a \num{450}$\times$ speed acceleration over the automated resonant 
platform proposed in \cite{janderson}, which requires \num{4} minutes to capture \num{35} data points. The speed test results conclusively highlight the superior processing efficiency and execution speed offered by the proposed hardware-software 
co-design, providing an ultra-fast solution for large-scale high-frequency core loss characterization.

\section{Conclusion}\label{sec5}
This paper presented a fully automated high-throughput core loss measurement system for sub-MHz power magnetics based on resonant methodologies. Analog conditioning circuits were utilized to identify fundamental resonance, and dual binary-scaled switched capacitor sequences were developed to enable automated frequency sweeping. The system architecture was optimized by selecting electromechanical power relays with great FOMs, thus minimizing the on-state resistance and the measurement errors. Extensive experimental evaluations including multiple magnetic materials within wide spectra were implemented to validate the feasibility, high fidelity, and exceptional speed of the automated measurement prototype, demonstrating close alignment with datasheets and unprecedented throughput with \num{1320} valid points within only \SI{20}{seconds}. The proposed hardware-firmware co-design provides a highly efficient solution for large-scale automated database generation, which is instrumental for the optimization of next-generation high-power-density power converters.

\vspace{12pt}
\color{red}

\end{document}